\documentclass{article}
\pdfoutput=1
\usepackage{color}
\usepackage{mathtools}
\usepackage{arxiv}
\usepackage{graphicx}
\usepackage{amsmath,tabu}
\usepackage{amsfonts}
\usepackage{amssymb}
\usepackage{makeidx}
\usepackage{graphicx}
\usepackage{lmodern}
\usepackage{float}
\usepackage[utf8]{inputenc} 
\usepackage[T1]{fontenc}    
\usepackage{hyperref}       
\usepackage{url}            
\usepackage{booktabs}       
\usepackage{amsfonts}       
\usepackage{nicefrac}       
\usepackage{microtype}      
\usepackage{lipsum}

\title{Investigating some attributes of periodicity in DNA sequences via semi-Markov modelling}

\author{
  Pavlos Kolias \\
  Department of Mathematics\\
              Aristotle University of Thessaloniki\\
              pakolias@math.auth.gr \\
   \And
 Alexandra Papadopoulou \\
           Department of Mathematics\\
              Aristotle University of Thessaloniki\\
              apapado@math.auth.gr \\
}

\begin{document}

\maketitle

\paragraph{Abstract}
DNA segments and sequences have been studied thoroughly during the past decades. One of the main problems in computational biology is the identification of exon-intron structures inside genes using mathematical techniques. Previous studies have used different methods, such as Fourier analysis and hidden-Markov models, in order to be able to predict which parts of a gene correspond to a protein encoding area. In this paper, a semi-Markov model is applied to 3-base periodic sequences, which characterize the protein-coding regions of the gene. Analytic forms of the related probabilities and the corresponding indexes are provided, which yield a description of the underlying periodic pattern. Last, the previous theoretical results are illustrated with DNA sequences of synthetic and real data.

\keywords{DNA sequences \and Periodicity \and Semi-Markov Chain}

\section{Introduction}
\paragraph{}
Periodicity is a structural property of DNA sequences. It is expressed as either nucleotides or words of nucleotides that appear with specific fixed distances in-between. Mainly, there have been observed two types of periodic behaviours in DNA. The first one was introduced by Trifonov in 1980 \cite{trifonov1980pitch} regarding chromatin, which is a basic element of the cell nucleus. Trifonov observed that certain di-nucleotides in the DNA of chromatin tends to appear at approximately every 10 to 11 bases. Subsequent studies suggested that the period of chromatin sequences converges to 10.4 bases \cite{cohanim2006specific}. Also, a more recent study \cite{salih2015visible}, which investigated the genome of three organisms, A. thaliana, C.elegans and H.sapiens, suggested that the di-nucleotide AA has almost perfect 10.5-base periodic behaviour in those organisms. One explanation about this type of periodicity is that the distance of 10.5 bases is exactly the "step" of the double strand, which curves the DNA chain and allows these long sequences to suppress into the  small area of the nucleus. The second type of periodicity has been observed in areas of the genome that are transcribed and later translated into proteins, also called coding regions. Previous studies have used methods from mathematical analysis, such as the spectral density, and they have shown that in coding regions, there is a tendency of certain nucleotides to reappear every 3-bases \cite{TSONIS1991323}. Also, this type of periodicity has only been observed in coding regions, while for non-coding regions there was not found any similar periodic behaviour. As each of the amino acids is encoded with a triplet of nucleotides (codons) and some specific amino acids are more abundant than others, authors concluded that the periodic behaviour, in fact exists, due to this higher frequency of certain amino acids and the period of 3-bases is sue to the triplet nature of the DNA. As the whole genome of each organism is frequently of several billions bases, the information about the periodic behaviour of the coding regions of the DNA would be really helpful into detecting those regions and distinguish between protein encoding regions and non-coding regions. Some algorithmic techniques have already been implemented using this information and they have used similar method, such as the Fourier transformation \cite{yin2007prediction}. Also, some other well-known algorithms use hidden-Markov models, in order to classify between different regions of DNA \cite{burge1997prediction}. In this paper we assume that a DNA sequence could be described by a semi-Markov chain $X_t$, with state space $S = \lbrace A,C,G,T \rbrace$, $t$ denotes the index position and $\boldsymbol{C}(m)= \lbrace c_{i,j}(m) \rbrace$ is the core matrix of the SMC. We propose a recursive formula based on the basic parameters of the model that could potentially identify regions that have "strong" or "weak" periodic behaviour. Finally we apply the model to both synthetic sequences and DNA sequences of several organisms.

\section{The semi-Markov model}
\paragraph{}
We assume that the DNA sequence is a realization of a semi-Markov chain $ X_n $ with state space the four nucleotides $ S = \lbrace A,C,G,T\rbrace $. The semi-Markov chain is described by a sequence of Markov transition matrices $ \lbrace \boldsymbol{P}(t) \rbrace_{t=0}^{\infty}$ and a sequence of conditional holding time matrices $\lbrace \boldsymbol{H}(m) \rbrace_{m=1}^{\infty}$, such as:
\begin{equation}
\boldsymbol{P}(t)= \lbrace p_{i,j}(t) \rbrace,
\end{equation}
where $p_{i,j}(t)=Prob[\text{the SMC will make its next transition to state j / the SMC entered state i at time t}]$
\vspace{0.2cm}
with $p_{i,j}(t)\geq 0 , \; \forall i,j\in S, \; t \in \mathbb{N} \quad$ and $\quad \sum_{j\in S}p_{i,j}(t) = 1, \; \forall i, \;  t \in \mathbb{N},$

and
\begin{equation}
\boldsymbol{H}(m)=\lbrace h_{i,j}(m) \rbrace,
\end{equation}
where $h_{i,j}(m) =Prob[\text{The SMC will stay in state i for $m$ positions before moving to state $j$}]$

\vspace{0.2cm}

We define the probabilities of the waiting time $w_{i}(m)$, which are the probabilities for the SMC to hold for $m$ time units in state $i$, before making its next transition.
\begin{equation}
w_{i}(m)= \sum_{j\in S}p_{i,j}h_{i,j}(m)
\end{equation}
Also the cumulative distribution for the waiting time is:
\begin{equation}
^{>}w_{i}(n)= \sum_{m=n+1}^{\infty}w_{i}(m)=\sum_{j\in S}p_{i,j}\:^{>}h_{i,j}(m)
\end{equation}

\vspace{0.2cm}

The basic parameter of the SMC is the \textit{core matrix} and it is defined as:
\begin{equation}
\boldsymbol{C}(m)= \lbrace c_{i,j}(t,m) \rbrace_{i,j \in S} = \boldsymbol{P}(t) \circ \boldsymbol{H}(m),
\end{equation}
where the operator $\lbrace \circ \rbrace$ denotes the element-wise product of matrices (Hadamard product).\\
We assume that DNA sequences do not contain virtual transitions, therefore: $p_{i,i}(t) = 0, \;\forall i\in S, \; t \in \mathbb{N} $. \\

\vspace{0.1cm}

We also define the interval transition probability $q_{i,j}(n)$, which is the probability for the SMC to be in state $j$ after $n$ time units, while it entered state $i$ in time $t=0$ to be: \cite{howard2012dynamic}
\begin{equation}
\boldsymbol{Q}(n)= \lbrace q_{i,j}(n) \rbrace_{i,j \in S} =\,^{>}\boldsymbol{W}(n)+\sum_{m=0}^{n}[\boldsymbol{P}\circ \boldsymbol{H}(m)]\boldsymbol{Q}(n-m),
\end{equation}
where $^{>}\boldsymbol{W}(n)=diag \lbrace \:^{>}w_{i}(n)\rbrace $.
\subsection{The homogeneous case}
\paragraph{}

\vspace{0.2cm}
In the following, the parameter of time is replaced by position, based on the nature of the DNA sequences, as their evolution depends on the index position of every letter in the sequence. In order to study the d-periodic behaviour of a DNA sequence, we would like to examine the probability of a letter appearance every $d$  steps. Thus, we define the following probability:
\begin{equation}
\begin{gathered}
p_{i}(d)=Prob[\text{the sequence will be in state $ i $, in position $d$}\\
\text{while it has been in state $i$ in the initial position}]
\end{gathered}
\end{equation}
 It is important to note that for a given DNA sequence, we do not know if the initial position is due to a letter transition or reappearance of the same letter, therefore we have to include both those two cases in calculating the above probability. We now present all the possible instances for the DNA sequence to be in state $i$, after $d$ steps, while it has been in state $i$ in the starting position.

 \vspace{0.2cm}
Let
\begin{equation}
S_x = \underset{x-times}{\underbrace{i\:i\:i \:i \: \cdots \: i}} \:j \: u \: u \cdots \:u \:i,
\end{equation}
the sequence of letters of length d, where $x = 1,2,...,d$, $j$ denotes any letter different than $i$ and $u$ denotes any letter from the state space $ S = \lbrace A,C,G,T\rbrace $

\begin{align*}
S_1 &= i\:j\:u \:u \: \cdots \:u \:i\\
S_2 &= i\:i\:j \:u \:u \cdots \:u\:i\\
S_3 &= i\:i\:i \:j\:u\:u \:\cdots \:u\:i\\
&\vdotswithin{=} \\
S_{d-2} &= i\:i\ \: \cdots \:i\:j\:u\:i \\
S_{d-1} &= \:i \:i\:i\ \: \cdots \:i\:j\:i \\
S_{d} &= \:i\:i\:i\:i\:i \: \cdots \:i\\
\end{align*}

\vspace{0.2cm}


The different instances $S_i$ are mutually exclusive and exhaustive events, thus using probabilistic argument we can conclude to the following equation, regarding the probability $p_i(d)$.
\begin{equation}
p_{i}(d)=\: ^{>}\!w_{i}(d)+ \sum_{j\neq i}^{N} \sum_{k=1}^{d}\,^{\geq}c_{i,j}(k)q_{j,i}(d-k)
\end{equation}

where $ ^{\geq}c_{i,j}(k)= p_{i,j} \cdot ^{\geq}\!h_{j,i}(k) \quad$, $^{\geq}\!h_{j,i}(k)$ denotes the survival function of the conditional holding times of the states and $q_{j,i}(d-k)$ is described in terms of the basic parameters of the semi-Markov chain and it follows
\begin{equation}
\begin{gathered}
q_{i,j}(n)=\delta_{i,j}\:^{>}w_{i}(n)+ \sum_{k\in S}p_{i,k}\:\sum_{m=0}^{n}h_{i,k}(m)q_{k,j}(n-m),\quad i,j\in S, \:n\in \mathbb{N},\\
\delta_{i,j} =\begin{cases}
 & 1  \quad i=j \\ 
 & 0  \quad i\neq j.
\end{cases}
\end{gathered}
\end{equation}

\vspace{0.2cm}

Equation 6 in matrix form is the following:
\begin{equation}
\boldsymbol{P}(d)=\: ^{>}\!\boldsymbol{W}(d)+\sum_{k=1}^{d}\, \boldsymbol{I} \circ \big[ [^{\geq}\boldsymbol{C}(k)\boldsymbol{Q}(d-k)][\boldsymbol{U-I}] \big]
\end{equation}
where $ \boldsymbol{U}\in M_{N\times N} $ is a square matrix with all the elements equal to 1 and $ ^{\geq}\!\boldsymbol{C}(k)= \boldsymbol{P} \circ \, ^{\geq}\!\boldsymbol{H}(k)$ and $\boldsymbol{Q}(n) = \lbrace q_{i,j}(n) \rbrace$.

\vspace{0.2cm}

For the interval transition probability matrix $\boldsymbol{Q}(n)$, instead of using the recursive formula, we can apply the closed analytic form, as proposed by Vassiliou and Papadopoulou \cite{vassiliou_papadopoulou_1992}.
\begin{equation}
 \begin{gathered}
	\boldsymbol{Q}(n) = \:^{>}\boldsymbol{W}(n) + \boldsymbol{C}(n)
+ \sum_{j=2}^{n} \lbrace \boldsymbol{C}(j-1)+\sum_{k=1}^{j-2}\boldsymbol{S}_j(k,m_k) \rbrace \\
\times \lbrace ^{>}\boldsymbol{W}(n-j+1)+ \boldsymbol{C}(n-j+1) \rbrace
\end{gathered}
\end{equation}
where
\footnotesize
\begin{equation}
\begin{gathered}
\boldsymbol{S}_j(k,m_k) = \sum_{m_k=2}^{j-k}\sum_{m_{k-1}=1+m_k}^{j-k+1} \cdots \sum_{m_1=1+m_2}^{j-1}\prod_{r=-1}^{k-1}\boldsymbol{C}(m_{k-r-1}-m_{k-r})
\end{gathered}
\end{equation}
\begin{center}
    for $j\geqslant k+2$, while if $j\leqslant k+2$ we have $\boldsymbol{S}_j(k,m_k)=0$.
\end{center}

\normalsize

\paragraph{}
For a "strongly" periodic chain, with period d, it is expected that for every periodic state, the frequency of the state appearances, every $ k \times d$ positions, would be high. So an interesting question is whether the chain is in the same state, not only for the first cycle of length $d$ but also for a number of $k$ successive cycles of the same length. Now, let $ \boldsymbol{P}(n,d)$ to be a column matrix with its i-th element to define the probability:
\begin{equation}
\begin{gathered}
p_{i}(n,d)=Prob[\text{the SMC to be in state i every d positions}\\ 
\text{for n cycles/ the initial state was $i$}]
\end{gathered}
\end{equation}
\vspace{0.2cm}
Using probabilistic argument and applying the equation 6, we can prove the following equation:
\begin{equation}
\boldsymbol{P}(n,d)=\boldsymbol{P}(n-1,d)\circ \Big[ ^{>}\!\boldsymbol{W}(d)+\sum_{k=1}^{d}\,  \boldsymbol{I} \circ [[^{\geq}\boldsymbol{C}(k)\boldsymbol{Q}(d-k)][\boldsymbol{U-I}] \Big]
\end{equation}
where $\boldsymbol{P}(n,d) = \lbrace p_i(n,d) \rbrace$. The initial condition is:

\begin{equation}
\boldsymbol{P}(1,d)=\: ^{>}\!\boldsymbol{W}(d)+\sum_{k=1}^{d}\, \boldsymbol{I} \circ [[^{\geq}\boldsymbol{C}(k)\boldsymbol{Q}(d-k)][\boldsymbol{U-I}]]
\end{equation}
Let us define the ratio $\boldsymbol{R}(n)$:

\begin{equation}
\boldsymbol{R}(n)= \big[ [\boldsymbol{P}(n-1,d)\boldsymbol{1}] \circ \boldsymbol{I} \big]^{-1} \cdot \boldsymbol{P}(n,d)
\end{equation}

where $\boldsymbol{1} = [ 1,1, ..., 1]$. The i-th element of matrix $\boldsymbol{R}(n)$ is the ratio of the probability $p_{i}(n,d)$ over $p_{i}(n-1,d)$ for every $n$ and illustrates the variations between the probabilities $p_{i}(n,d)$ and $p_{i}(n-1,d)$, in order to investigate the periodicity over a number of cycles.
\vspace{0.2cm}

\vspace{0.2cm}

\subsection{The case of partial non homogeneity}
\paragraph{}

\vspace{0.2cm}

The partial non-homogeneous semi-Markov chain is constructed based on the fact that every amino acid consists of three nucleotides (codon). Using this information we can create three discrete coding positions $k = \lbrace 1,2,3\rbrace $ and for the NHSMC we have three stochastic matrices $\boldsymbol{P}(k),\; k = 1,2,3$ for the embedded Markov chain. In order to investigate the periodic behaviour, we define the following probability:
\vspace{0.2cm}
\begin{equation}
\begin{gathered}
p_{i}(k,d)=Prob[ \text{\small{the NHSMC will be in state $i$ in the position $d$}}\\
\text{\small{/ initially the NHSMC was in state $i$ in coding position $k$}}]
\end{gathered}
\end{equation}

We now present all the possible and mutually exclusive events for the realization of the event of the probability $p_{i}(k,d)$.
\vspace{0.2cm}
\footnotesize
\begin{align*}
S_1 &= i(k) \: j(k+1) \:u(k+2) \:u(k+3) \:u (k+4) \:  \cdots \:u((k+d-1) \bmod s) \:i((k+d) \bmod s)\\
S_2 &= i(k) \:i(k+1) \:j(k+2)  \:u(k+3)  \:u(k+4)  \cdots \:u((k+d-1)\bmod s) \:i((k+d)\bmod s) \\
S_3 &= i(k) \:i(k+1)\:i(k+2) \:j(k+3)\:u(k+4)\:\cdots \:u((k+d-1)\bmod s) \:i((k+d)\bmod s)\\
    &\vdotswithin{=} \\
S_{d-2} &= i(k)\:i(k+1)\:i(k+2) \: \cdots \:j((k+d-2)\bmod s) \:u((k+d-1)\bmod s)\:i((k+d)\bmod s) \\
S_{d-1} &= \:i(k) \:i(k+1)\:i(k+2)\:i(k+3) \: \cdots \:j((k+d-1)\bmod s)\:i((k+d)\bmod s) \\
S_{d} &= \:i(k)\:i(k+1)\:i(k+2)\:i(k+3)\:i(k+4)\:i(k+5) \: \cdots \:i((k+d)\bmod s)\\,
\end{align*}
\normalsize
where $j(\cdot)\neq i(\cdot)$ and $u(\cdot)$ denotes a letter from the state space S.
\vspace{0.2cm}

It is easy to show that the different $ S_i $ events are mutually exclusive and cover the whole sample space, thus we can conclude to the following equation for the probability $p_{i}(k,d)$
\begin{equation}
p_{i}(k,d)=^{>}\!w_(k,d)+ \sum_{j\neq i}^{N} \sum_{x=1}^{d}\,^{\geq}c_{i,j}(k,x)q_{j,i}((k+x) \bmod s,d-x)
\end{equation}

The quantities $^{>}\!w_{i}(\cdot), \quad ^{\geq}c_{i,j}(\cdot),$ and $\quad q_{j,i}(\cdot) $ are functions of the basic parameters $\boldsymbol{P}(k)$ and $\boldsymbol{H}(m)$ of the NHSMC. The interval transition probabilities $q_{j,i}(\cdot)$ are expressed by the following equation:

\begin{equation}
q_{i,j}(k,n)=\delta_{i,j}\:^{>}w_{i}(k,n)+ \sum_{x\in S}p_{i,x}(k)\:\sum_{m=0}^{n}h_{i,x}(m)q_{x,j}(k+m,n-m),\quad i,j\in S, \:n\in \mathbb{N}
\end{equation}

\vspace{0.2cm}

where ${>}w_{i}(n,s)$ denotes the survival function of the unconditional holding times for the state $i$.

\vspace{0.2cm}

Using matrix notation we can write the equation 16 as:
\begin{equation}
\boldsymbol{P}(k,d)=^{>}\!\boldsymbol{W}(k,d)+ \sum_{x=1}^{d}\, \boldsymbol{I}\circ [[^{\geq}\boldsymbol{C}(k,x)\boldsymbol{Q}((k+x) \bmod s,d-x)][\boldsymbol{U-I}]]
\end{equation}
The elements of the matrix $\boldsymbol{Q}((k+x) \bmod s,d-x)$ are the interval transition probabilities for the NHSMC, which could be expressed by the following recursive formula:
\begin{equation}
\boldsymbol{Q}(s,n)=\:^{>}\boldsymbol{W}(s,n)+ \sum_{m=1}^{n}\boldsymbol{C}(s,m)\boldsymbol{Q}(s+m,n-m)
\end{equation}

\vspace{0.2cm}

For the recursive equation of the interval transition probabilities for the NHSMC (19), we also have the closed analytic form \cite{vassiliou_papadopoulou_1992}:
\begin{equation}
	\begin{gathered}
	\boldsymbol{Q}(k,n) = \:^{>}\boldsymbol{W}(k,n) + \boldsymbol{C}(k,n)
+ \sum_{j=2}^{n} \lbrace \boldsymbol{C}(k,j-1)+\sum_{x=1}^{j-2}\boldsymbol{S}_j(x,k,m_x) \rbrace \\
\times \lbrace ^{>}\boldsymbol{W}(k+j-1,n-j+1)+ \boldsymbol{C}(k+j-1,n-j+1) \rbrace
\end{gathered}
\end{equation}
where
\footnotesize
\begin{equation}
\begin{gathered}
\boldsymbol{S}_j(x,k,m_x) = \sum_{m_x=2}^{j-x}\sum_{m_{x-1}=1+m_x}^{j-x+1} \cdots \sum_{m_1=1+m_2}^{j-1}\prod_{r=-1}^{x-1}\boldsymbol{C}(k+m_{x-r}-1,m_{x-r-1}-m_{x-r})
\end{gathered}
\end{equation}
for $j\geqslant x+2$, while if $j\leqslant x+2$ we have $\boldsymbol{S}_j(x,k,m_x)=0$.

\vspace{0.2cm}
\normalsize

Similarly with the homogeneous case, we are interested for the sequence to be in the same state, not only after $d$ steps, but also for a number n of successive cycles of length d, given that its initial coding position was $k$. Let $\boldsymbol{P}(k,n,d)$ to be a column matrix and its i-th element to define the probability:
\begin{equation}
\begin{gathered}
p_{i}(k,n,d)=Prob[\text{the NHSMC will be in state $i$ every $d$ positions for}\\ 
\text{n cycles / the initial state was $i$ in coding position $k$}]
\end{gathered}
\end{equation}
Using probabilistic argument and the equation 18, we can prove the following equation:
\begin{equation}
\begin{gathered}
\boldsymbol{P}(k,n,d)=\boldsymbol{P}(k,n-1,d)\circ \\
\Big[ ^{>}\!\boldsymbol{W}(k,d)+\sum_{x=1}^{d}\, \boldsymbol{I}\circ [[^{\geq}\boldsymbol{C}(k,x)\boldsymbol{Q}((k+x) \bmod s,d-x)][\boldsymbol{U-I}]] \Big]
\end{gathered}
\end{equation}
where $\boldsymbol{U}= \lbrace u_{i,j} \rbrace_{i,j \in S}, \quad u_{i,j}=1, \quad \forall i,j \quad$, $^{>}\!\boldsymbol{W}(k,d) = diag\lbrace ^{>}w_i(k,d) \rbrace \quad $,  $^{\geq}\!\boldsymbol{C}(k,m)= \boldsymbol{P}(k) \circ \, ^{\geq}\!\boldsymbol{H}(m)$ and $\boldsymbol{Q}(k,n) = \lbrace q_{i,j}(k,n)_{i,j \in S} \rbrace$.
\vspace{0.2cm}
The initial condition is:

\begin{equation}
\boldsymbol{P}(k,1,d)=\: ^{>}\!\boldsymbol{W}(d)+\sum_{k=1}^{d}\, \boldsymbol{I} \circ [[^{\geq}\boldsymbol{C}(k)\boldsymbol{Q}(d-k)][\boldsymbol{U-I}]]
\end{equation}
We define the ratio $\boldsymbol{R}(k,n)$:

\begin{equation}
\boldsymbol{R}(k,n)= \big[ [\boldsymbol{P}(k,n-1,d)\boldsymbol{1}] \circ \boldsymbol{I} \big]^{-1} \cdot \boldsymbol{P}(k,n,d)
\end{equation}

where $\boldsymbol{1} = [ 1,1, ..., 1]$. The i-th element of matrix $\boldsymbol{R}(k,n)$ is the ratio of the probability $p_{i}(k,n,d)$ over $p_{i}(k,n-1,d)$ for every $n$ and illustrates the variations between the probabilities $p_{i}(k,n,d)$ and $p_{i}(k,n-1,d)$, in order to investigate the periodicity over a number of cycles, with a specific coding position $k$.

\section{Illustrations of real and synthetic data}
\paragraph{}
For the illustrations of the homogeneous semi-Markov model, synthetic DNA sequences as well as real genomic and mRNA sequences were used. The coding sequence was human dystrophin mRNA and the non-coding sequence, which was used for comparison, was the human b-nerve growth factor gene (BNGF). We assumed that each of the sequences could be described by a homogeneous semi-Markov chain $\lbrace X_t \rbrace _{t=0}^{\infty}$, with state space $S = \lbrace A,C,G,T\rbrace$ and the index $t$ denotes the position of each nucleotide inside the sequence. The basic parameters $\boldsymbol{P}_{i,j}(s)$ and $\boldsymbol{H}_{i,j}(m)$ of the SMC were estimated using the empirical estimators:

\begin{equation}
\widehat{p}_{i,j}(k) = \frac{N(i(k)\rightarrow j)}{\sum_{x\in S}N(i(k)\rightarrow x)} \qquad \text{and} \qquad \widehat{h}_{i,j}(m) = \frac{N(i\rightarrow j,m)}{\sum_{x\in S}N(i\rightarrow x,m)},
\end{equation}
where $N(i(k)\rightarrow j)$ denotes the number of transitions from state $i$ to state $j$, starting from coding position $k$ and $N(i\rightarrow j,m)$ denotes the number of transitions from state $i$ to state $j$, while the SMC remained in state $i$ for $m$ positions.
\paragraph{}
In order to estimate the initial condition, which are the probabilities of the matrix $\boldsymbol{P}(k,1,d)$, the first 10 cycles of length 3 have been used and the basic parameters $\boldsymbol{P}(k)$ and $\boldsymbol{H}(m)$ have been estimated. After that and for each cycle $n$, the core matrix has been estimated $\boldsymbol{C}(k,m)$, using the letters of the sequence up until the position $n \cdot d + k$. This specific process has been implemented, correcting the estimations, as in the current application the length of each period is small ($d=3$), resulting  in an non adequate sample size for each cycle. Finally, the probability for the chain to be in the same state for every $n\cdot d$ positions has been calculated using:
\begin{equation}
\begin{gathered}
\boldsymbol{P}(k,n,d)=\boldsymbol{P}(k,n-1,d)\circ \\
\Big[ ^{>}\!\boldsymbol{W}(k,d)+\sum_{x=1}^{d}\, \boldsymbol{I}\circ [[^{\geq}\boldsymbol{C}(k,x)\boldsymbol{Q}((k+x) \bmod s,d-x)][\boldsymbol{U-I}]] \Big]
\end{gathered}
\end{equation}

\subsection{DNA sequences of synthetic data}

\paragraph{Example 1: Comparison between random and periodic DNA sequences}\mbox{}\\

\vspace{0.05cm}

Let $L$ a DNA sequence of length $N=1000$ of the form: $L=\lbrace U,U,U,U,U,U,U,U,U,U,U,U,U...\rbrace$, where the letter $U$ corresponds to any nucleotide, from a uniform distribution\\
\begin{center}
$Prob[U = A]=Prob[U=C]=Prob[U=G]=Prob[U=T]= \frac{1}{4}$.
\end{center}
This kind of sequence would not exhibit any periodic behaviour, however the estimated probability matrix $\boldsymbol{P}(n,d)$ for $d=3$ will be estimated for comparison. The estimation of the embedded Markov matrix $\boldsymbol{P}$ is:
\begin{center}
$\boldsymbol{P}= \begin{pmatrix}
0 & 0.2 & 0.8 & 0 \\
0.375 & 0 & 0.5 & 0.125 \\
0.125 & 0.5 & 0 & 0.375 \\
0.25 & 0.75 & 0 & 0 \\
\end{pmatrix}$
\end{center}
and the core matrix $\boldsymbol{C}(m)$ is:
\begin{center}
$\boldsymbol{C(1)}= \begin{pmatrix}
0 & 0 & 0.8 & 0 \\
0.375 & 0 & 0.5 & 0.125 \\
0.125 & 0.375 & 0 & 0.375 \\
0.25 & 0.5 & 0 & 0 \\
\end{pmatrix}$
\end{center}
and
\begin{center}
$\boldsymbol{C(2)}= \begin{pmatrix}
0 & 0 & 0 & 0 \\
0 & 0 & 0 & 0 \\
0 & 0.125 & 0 & 0 \\
0 & 0 & 0 & 0 \\
\end{pmatrix}$,
\end{center}
while the only non zero element of  $\boldsymbol{C}(3)$ is $c_{4,2}(3) = 0.25$
The initial condition $\boldsymbol{P}(1,3)$ is:
\begin{center}
$\boldsymbol{P}(1,3)= \begin{pmatrix}
0.32\\
0.34\\
0.42\\
0.27\\
\end{pmatrix}$
\end{center}
\begin{figure}[H]
\centering
    \includegraphics[trim={0 0 0 2cm},clip,keepaspectratio, scale=0.8]{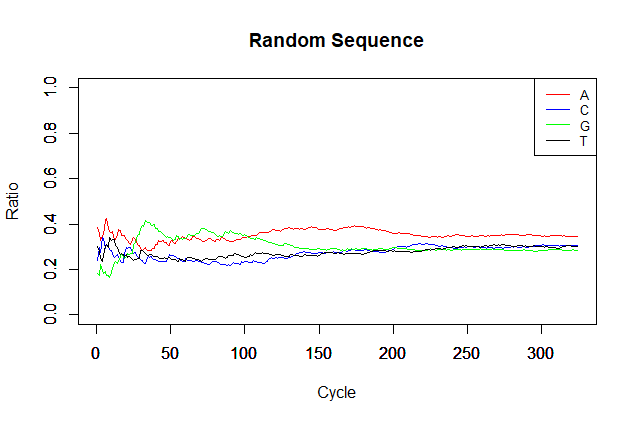}
    \caption{R(n) for the synthetic DNA sequence of a uniform distribution}
\end{figure}

\paragraph{}

Now let $L$ a DNA sequence of length $N=1000$ of the form: $L=\lbrace A,U,U,A,U,U,A,U,U,A,U,U,A...\rbrace$, where the letter $A$ corresponds to adenine, while the letter $U$ corresponds to any nucleotide from the uniform distribution\\
\begin{center}
$Prob[U = A]=Prob[U=C]=Prob[U=G]=Prob[U=T]= \frac{1}{4}$.
\end{center}
We will investigate the periodic behaviour, of period $d=3$. One can notice that for the letter  $A$ can have a waiting time $w_A{m}$ for every $m$. Pn the other hand, for the other three letters $C,G,T$, the waiting times are zero if $m$ exceeds two, as between 3 letters, there always exists the letter $A$. The estimated embedded Markov matrix $\boldsymbol{P}$ is:
\begin{center}
$\boldsymbol{P} = \begin{pmatrix}
0 & 0.30 & 0.30 & 0.40 \\
0.73 & 0 & 0.15 & 0.12 \\
0.69 & 0.17 & 0 & 0.14 \\
0.70 & 0.14 & 0.16 & 0 \\
\end{pmatrix}$
\end{center}
and the core matrix is:
\begin{center}
$\boldsymbol{C}(1)= \begin{pmatrix}
0 & 0.19 & 0.16 & 0.27 \\
0.60 & 0 & 0.15 & 0.13 \\
0.56 & 0.17 & 0 & 0.15 \\
0.50 & 0.14 & 0.16 & 0 \\
\end{pmatrix}$
\end{center}
and
\begin{center}
$\boldsymbol{C}(2)= \begin{pmatrix}
0 & 0.08 & 0.11 & 0.09 \\
0.13 & 0 & 0 & 0 \\
0.13 & 0 & 0 & 0 \\
0.20 & 0 & 0 & 0 \\
\end{pmatrix}$
\end{center}
while the other matrices $\boldsymbol{C}(m)$ for $m>2$ have non zero elements only in the first row. The initial condition $\boldsymbol{P}(1,3)$ is:
\begin{center}
$\boldsymbol{P}(1,3)= \begin{pmatrix}
0.83\\
0.18\\
0.20\\
0.25\\
\end{pmatrix}$
\end{center}
The probability for the chain to be in state $A$, every  $d=3$ positions, while starting from state $A$, is greater than the other three states, as we expected. However, the probability $p_{A}(n,3)$ is lower than 1, because it is also allowed for the SMC to be in state $A$ in-between a periodic cycle.
\begin{figure}[H]
\centering
    \includegraphics[trim={0 0 0 2cm},clip,keepaspectratio, scale=0.8]{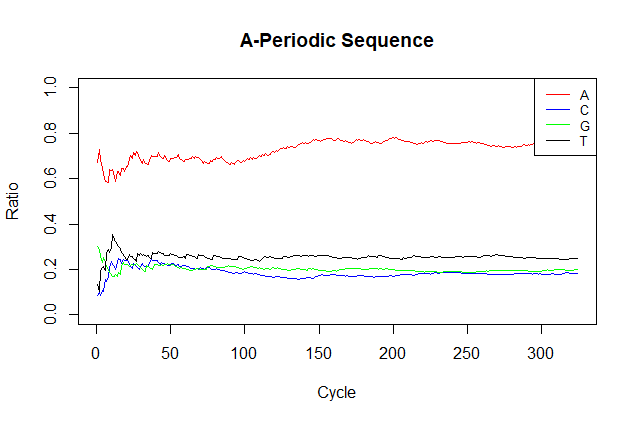}
    \caption{R(n) for the synthetic DNA sequence with 3-base periodicity of adenine}
\end{figure}
\paragraph{Example 2: Detection of periodic regions inside a sequence}\mbox{}\\

Let $L$ a DNA sequence of length $N=5000$ of the form: $L=\lbrace U,U,U,U,U,U,U,U,U,U,U,U,U...\rbrace$, where the letter $U$ corresponds to any random nucleotide. In the intervals $1500-2000$ and $3000-3500$, which correspond to the cycles $500-666$ and $1000-1166$, the letter $U$ has been substituted with the letter $A$, starting from the first position of each interval and for every 3 positions. Figure 1 shows the values of the ratio $R(n)$ for the letter $A$, where the green regions are the cycles of the sequence $R(n)$ where the sequence is increasing, while the red regions are the cycles where the sequence $R(n)$ decreases. It is observed from the figure, that the regions, in which we have synthetically added periodic behaviour are apparently colored with green.
\begin{figure}[H]
\centering
    \includegraphics[trim={0 0 0 2cm},clip,keepaspectratio, scale=0.8]{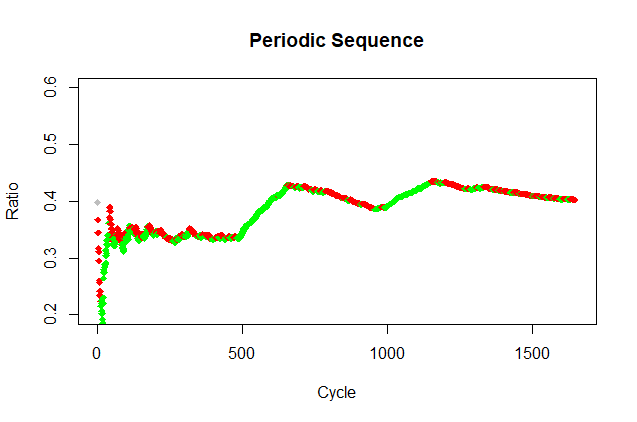}
    \caption{R(n) of the letter $A$ of the synthetic sequence with periodicity in the cycles 500-666 and 1000-1166}
\end{figure}

\subsection{DNA sequences of real data}
The information about the periodic behaviour of the coding regions of the genome could possibly be used, in order to distinguish these regions, over a DNA sequence with great length. For the coding sequences of real DNA, the human dystrophin mRNA has been used, while for the non coding region, the human b-nerve growth factor has been used. These sequences have a length greater than 5000 bases and they have already been studied for periodic behaviour \cite{TSONIS1991323}
\begin{figure}[H]
\centering
\includegraphics[trim={0 0 0 1.5cm},clip,scale=0.8]{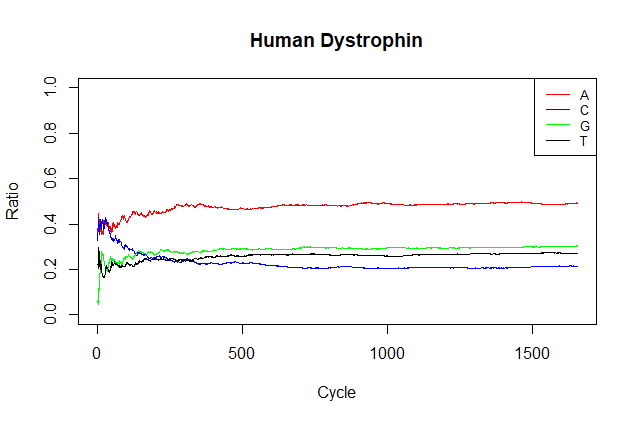}
\caption{$R(n)$ for the human dystrophin mRNA sequence}
\end{figure}

\begin{figure}[H]
\centering
\includegraphics[trim={0 0 0 2cm},clip,scale=0.8]{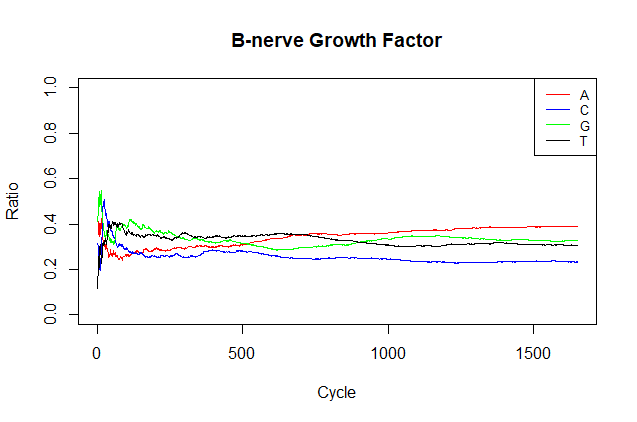}
\caption{$R(n)$ for the human b-nerve growth factor sequence}
\end{figure}
It is obvious that the probabilities $p_{i}(k,n,d)$ will converge to zero, as they are a product of n probabilities. The most important things in the investigation are the initial probability $\boldsymbol{P}(k,1,d)$, which contains the probabilities for the chain to be in the same state after $d$ positions and also the ratio $\boldsymbol{R}(k,n)$, which measures the relationship between the probabilities of the current cycle and the previous one. If the values of $\boldsymbol{R}(k,n)$ are high, then the probabilities $p_{i}(k,n,d)$ decrease with a slow rate, while if the values of $\boldsymbol{R}(k,n)$ are low, then the probabilities $p_{i}(k,n,d)$ decrease with a slow rate. One can notice that for the human dystrophin mRNA sequence, the nucleotide $A$ has a higher chance to appear every $3$ positions, while all the other nucleotides have the same behaviour.On the other hand, for the human b-nerve growth factor, all the nucleotides have approximately the same probability to appear every $3$ positions.

\section{Conclusion}
\paragraph{}

In the present paper, a method was developed, in order to investigate the periodicity of DNA sequences. The model was developed using a semi-Markov chain and the basic parameters were calculated using recursive equations for a number of cycles of a specified length. The idea for the development of this method occurred by a main problem in computational biology, that is the  identification of coding and non-coding regions over a long DNA sequence. From previous studies, it is known that the coding regions of the gene have different structure from the non-coding regions, as they exhibit a characteristic tendency of repetition of some nucleotides every 3 positions. Using this fact and by modelling a DNA sequence as a semi-Markov chain, the probabilities of the chain to be in the same state every $ d $ positions for the entire length, were calculated. The numerical results of the implementation of the model on actual data confirmed the previous studies, as it was apparent that periodic behaviour is a characteristic of the coding segments, unlike non-coding segment that did not show similar behaviour. For the estimation of the parameters, a correction procedure was applied, due to the short duration of the period ($ d = 3 $) for the specific application. The algorithm could potentially be used as an initial method for investigating periodicity for any DNA sequence and also it could be used to separate two different DNA segments in terms of periodic behaviour. Although the examples produced satisfactory results, they should be perceived with caution, due to the complexity of the structure of DNA and its various peculiarities. For example, additional parameters could be included in the model, such as the sequence length, the frequencies of each nucleotide, the open reading frames (orf's), the species of the organism, the mutations, and others. Also, because in DNA sequences the characteristic of periodicity still exists, even when there are small perturbations in the cycle of the period, such as a shift of the position of a letter, an interesting question for general modelling, would be to study this specific problem under this case.
\bibliographystyle{amsplain}
\bibliography{References}

\end{document}